\documentclass{sig-alternate-05-2015}

\usepackage{etoolbox}
\makeatletter
\patchcmd{\maketitle}{\@copyrightspace}{}{}{}
\makeatother

\usepackage{soul}
\usepackage{pifont}
\usepackage{graphicx}
\usepackage{amsmath}
\usepackage{array}
\usepackage{mybeamer}
\newcommand{\nonl}{\renewcommand{\nl}{\let\nl\oldnl}}
\newcommand{\eg}{\mbox{{\em e.g.}}}

\makeatletter
\newcommand{\removelatexerror}{\let\@latex@error\@gobble}
\makeatother

\usepackage{xcolor}

\begin{document}






%


\title{A Post-Silicon Trace Analysis Approach for System-on-Chip Protocol Debug}

%
%
%
%
%

\numberofauthors{2} 
%
\author{
%
%
\alignauthor
Yuting Cao, Hao Zheng\\
       \affaddr{Computer Science \& Engineering}\\
       \affaddr{U. South Florida}\\
       \affaddr{Tampa, FL 33620}\\
       \email{\{cao2, haozheng\}@usf.edu}
\alignauthor
Sandip Ray, Jin Yang\\
	    \affaddr{Strategic CAD Lab}\\
       \affaddr{Intel}\\
       \affaddr{Hillsboro, OR}\\
       \email{\{sandip.ray, jin.yang\}@intel.com}
}
\def\checkmark{\tikz\fill[scale=0.4](0,.35) -- (.25,0) -- (1,.7) -- (.25,.15) -- cycle;} 
\maketitle
\begin{abstract}
Reconstructing system-level behavior from silicon traces is a critical problem in post-silicon validation of System-on-Chip designs.  Current industrial practice in this area is primarily manual, depending on collaborative insights of the architects, designers, and validators.  This paper presents a trace analysis approach that exploits architectural models of the system-level protocols to reconstruct design behavior from partially observed silicon traces in the presence of ambiguous and noisy data.  The output of the approach is a set of all potential interpretations of a system's internal executions abstracted to the system-level protocols.  To support the trace analysis approach, a companion trace signal selection framework guided by system-level protocols is also presented, and its impacts on the complexity and accuracy of the analysis approach are discussed. That approach and the framework have been evaluated on a multi-core system-on-chip prototype that implements a set of common industrial system-level protocols.  
\end{abstract}

%
%

%
%

%
%



\section{Introduction}

Post-silicon validation makes use of pre-production silicon
integrated circuit (IC) to ensure that the fabricated system
works as desired under actual operating conditions with real
software. 
It is a critical component of the design validation life-cycle for modern microprocessors and system-on-chip (SoC) designs.  
Unfortunately, it is also highly complex, performed under aggressive schedules and accounting for more than $50\%$ of the overall design validation cost~\cite{Patra2007}. 

An SoC design is often composed of a large number of pre-designed hardware or
software blocks (often referred to as ``intellectual
properties'' or ``IPs'') that coordinate through complex
protocols to implement system-level behavior~\cite{Foster2015DAC}. 
An execution trace of a system typically involves activities from
the CPU, audio controller, display controller, wireless
radio antenna, etc., reflecting the interleaved execution of a potentially large number of 
communication protocols. 
As SoCs integrate more IPs,
the interactions among the IPs are increasingly more complex. 
Moreover, modern interconnects are highly concurrent allowing
multiple transactions to be processed simultaneously for scalability
and performance. They are an important source of design errors.
On the other hand, observability limitations allow only a small number of
participating signals to be actually traced during silicon
execution.  Furthermore, electrical perturbations cause
silicon data to be noisy, lossy, and ambiguous. It is non-trivial during post-silicon debug 
to identify all participating protocols and pinpoint the interleavings that
result in an observed trace.  

Previous work~\cite{Zheng2016ISQED} proposed a method for correlating silicon traces with system-level protocol specifications.  The idea was to reconstruct protocol execution scenarios from a partially observed silicon trace, which provide abstract views of system internal executions to facilitate post-silicon SoC debug.  While that work showed promising results, it has a number of deficiencies precluding its applicability in practice.  First, there was no way to qualify or rank the quality of protocol execution scenarios generated by the reconstruction procedure.  Under poor observability condition, it was possible for the algorithm to generate hundreds or thousands of potential protocol execution scenarios consistent with a partially observed trace. Without a metric to rank the quality of these reconstructions, the debugger is faced with the unenviable task of wading through these potential scenarios to infer what may actually have happened in a specific silicon execution.  
Moreover, based on past experiences, interleavings of different protocol executions are a major source of functional bugs.  Since the method developed in~\cite{Zheng2016ISQED} does not capture orderings among different protocol executions, the results obtained with that method offer little help for bug localization and root causing. 

This paper addresses the above deficiencies by introducing an optimized trace analysis approach.  Central to this optimized approach is a new formulation of protocol execution scenarios that comprehends ordering relations among protocol executions.  Quantitative metrics are also developed so that the quality of the results derived by and the efficiency of the analysis approach can be measured.  Trace signal selections can have great impacts on the complexity and accuracy of the trace analysis.  Therefore, a companion trace signal selection framework is proposed.  This framework is communication-centric, and guided by system-level protocols.  {\em Its objective is to facilitate the trace analysis to produce high quality interpretations of observed silicon traces efficiently}.  Various trace signal selection strategies are evaluated and analyzed based on their impacts on the trace analysis approach applied to a non-trivial multi-core SoC model that implements a number of common industrial system-level protocols.


%
%
\section{Flow Specification}

\begin{figure}[tb]
\begin{center}
\resizebox{2.7in}{!}{
\includegraphics[width=3.3in]{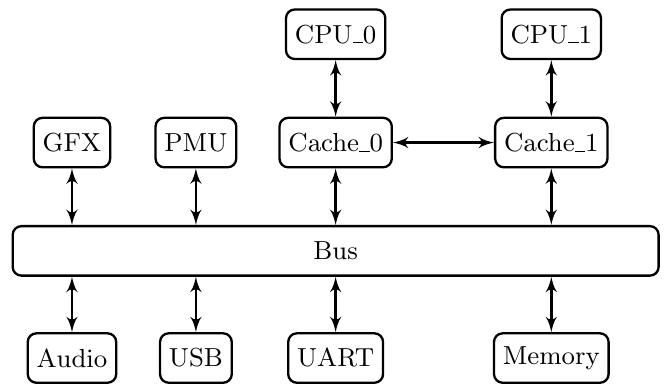}
}
\end{center}
\caption{The block diagram of the simple SoC model.}
\label{rtlstruc}
\end{figure}

\begin{figure}[tb]
\begin{center}
\resizebox{.4\textwidth}{!}{
\includegraphics[width=2in]{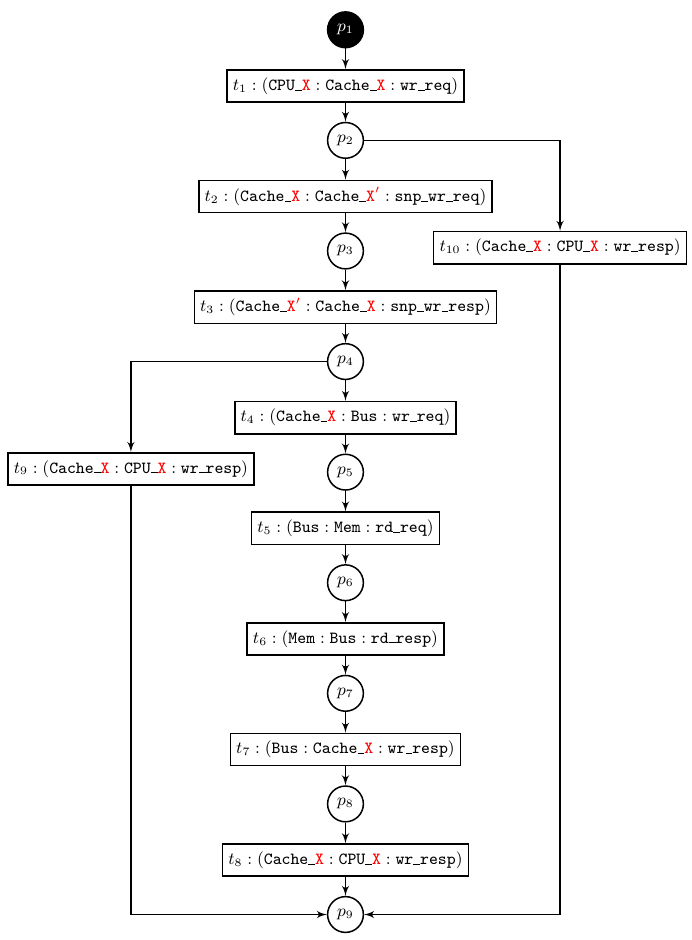}
}
\caption{LPN formalization of a CPU write protocol. Each LPN transition is labeled with an event $({\tt src, dest, cmd})$ where ${\tt cmd}$ is a command sent from a source
  component ${\tt src}$ to a destination component ${\tt dest}$. The places without outgoing edges are \emph{terminals}, which indicate termination of protocols represented by the LPNs. 
}
\label{fig:flow-spec-ex}
\end{center}
\end{figure}

An SoC model as shown in Figure~\ref{rtlstruc} is used to illustrate and experiment the work described in this paper.  It consists of two CPUs (CPU\_X), each with a private {Data Cache} (Cache\_X), a graphics engine (GFX), a power management unit (PMU),  a system memory, and three peripheral blocks: an audio control unit ({Audio}), a UART controller ({UART}), and a USB controller ({USB}).  All these blocks are connected through an interconnect fabric ({Bus}).

System operations are realized by executions performed in various blocks that are coordinated by system-level protocols.  These protocols are typically specified in architecture documents as message flow diagrams, where the words ``protocol'' and ``flow'' are used interchangeably.  In this paper, as in~\cite{Zheng2016ISQED}, system flows are formalized using Labeled Petri-nets (LPNs).  Figure~\ref{fig:flow-spec-ex} shows a memory write protocol initiated from a CPU \texttt{CPU\_X} in LPN where ${\tt X} \in \{0,1\}$ and ${\tt X' = 1 - X}$.
An LPN is a tuple {$(P, T, E, L, s_0)$} where {$P$} is
a finite set of \emph{places}, $T$ is a finite set of \emph{transitions}, $E$ is a finite set of \emph{events}, and $L: T \rightarrow E$ is a labeling function that maps each
transition $t \in T$ to an event $e \in E$.  For each
transition $t \in T$, its preset, denoted as $\bullet{t}
\subseteq P$, is the set of places connected to $t$, and its
postset, denoted as $t\bullet \subseteq P$, is the set of
places that $t$ is connected to.  A state $s \subseteq P$
of a LPN is a subset of places marked with tokens.  There are two special states associated with each LPN; $s_0 \subseteq P$ which is the set of initially
marked places, also referred to as the \emph{initial state}, and the end state $s_{end}$ which is the set of places not going to any transitions.  

A transition $t$ can be executed in a state $s$ if $\bullet t \subseteq s$.  Executing $t$ causes the labeled event to be emitted, and leads to a new state $s^\prime = (s - \bullet t) \cup t \bullet$. Therefore, executing an LPN leads to a sequence of events.  Execution of a LPN completes if its $s_{end}$ is reached.  

For example, in Figure~\ref{fig:flow-spec-ex}, $t_1$ can be executed in $s_0 = \{p_1\}$.  Event ${\tt (CPU\_X:Cache\_X:wr\_req)}$ is emitted after $t_1$ is executed, and the LPN state becomes $\{p_2\}$.  The  end state is $s_{end} =\{p_{9}\}$. 

A flow specification may also contain multiple branches describing different ways a system can execute such flow. For example, the flow shown in Figure~\ref{fig:flow-spec-ex} has three branches covering the cases where the cache (snoop) operation is hit or miss.


\section{Post-silicon Trace Analysis}

\subsection{Previous Work}
\label{sec:previous-work}

This section recaps the previous approach in~\cite{Zheng2016ISQED}.  The objective of the trace analysis is to reconstruct design internal behavior {\em wrt}  given system-level flow specifications $\mathbf{F}$ from a partially observed silicon trace on a small number of hardware signals.  The off-chip analysis includes two broad phases: (1)~trace abstraction, which maps a silicon  trace into a sequence of \emph{flow events}, higher-level architectural constructs including  \eg, messages, operations, etc, and (2)~trace interpretation, which infers possible flow execution scenarios that are compliant with the abstracted event sequence.  

 


To illustrate the basic idea, consider the system flow in
Figure~\ref{fig:flow-spec-ex}, which we call $F_1$.
Suppose that the following flow execution trace is abstracted from an
observed silicon trace by executing a design that implements $F_1$.
\[
	t_1\;t_2\;t_1\;t_2\;t_3\;t_3\;t_4\;t_4\;t_5\;t_6\;t_5\;t_6\ldots
\]  
Here the flow events are referred to by their transition names in the LPN.  The first four events result in
the following flow execution scenario
\begin{equation}
\label{ex:analysis:1}
	\{(F_{1,1}, \{p_3\}),~(F_{1,2}, \{p_3\})\}.
\end{equation}
A flow execution scenario is defined as a set of flow instances and their respective current states after some events are processed~\cite{Zheng2016ISQED}.  It can be viewed as an abstraction of system states \textit{wrt} system flows.  The above execution scenario indicates that the sequence of the first four events is a result from executing those two flow instances of $F_1$ from their initial states to the shown states.   For the first event $t_3$, it may be a result from executing $F_{1,1}$ or $F_{1,2}$, but exactly which one is unknown due to limited observability.  Both possible cases are considered, and  two execution scenarios below are derived as a result from interpreting $t_3$.
\begin{equation}
\label{ex:analysis:2}
\begin{array}{l}
	\{(F_{1,1}, \{p_4\}),~(F_{1,2}, \{p_3\})\} \\
	\{(F_{1,1}, \{p_3\}),~(F_{1,2}, \{p_4\})\}.
\end{array}
\end{equation}
After handling the next event $t_3$, the above two execution scenarios
are reduced to the one as shown below.
\[
	\{(F_{1,1}, \{p_4\}),~(F_{1,2}, \{p_4\})\}.
\]
After the remaining six events are handled, the following execution scenario is derived.
\[
	\{(F_{1,1}, \{p_{7}\}),~(F_{1,2}, \{p_{7}\})\}
\]

As another example, now suppose that the design with a bug generates the flow trace below.
\[
	t_1\;t_2\;t_1\;t_2\;t_3\;t_3\;t_4\;t_4\;t_5\;t_6\;t_5\;t_{11}\ldots.
\] 
This sequence is almost the same as the previous one except that the last event is $t_{11}: \mbox{\tt (Cache\_X:CPU\_X:rd\_resp)}$ instead of $t_6: \mbox{\tt (Mem:Bus:rd\_resp)}$ in the previous trace.  $t_{11}$ is an event used in a different flow specification describing a CPU memory read protocol.  Analyzing the trace right before $t_{11}$ leads to the execution scenarios below.
\begin{equation}
\label{ex:analysis:3}
\begin{array}{l}
	\{(F_{1,1}, \{p_{7}\}),~(F_{1,2}, \{p_{6}\})\} \\
	\{(F_{1,1}, \{p_{6}\}),~(F_{1,2}, \{p_{7}\})\}.
\end{array}
\end{equation}
However, $t_{11}$ cannot be a result from executing either flow instances in both scenarios, which indicates a noncompliance of the design implementation with respect to the given flow specification.  Such an event is referred to as being \emph{inconsistent}.  In this case, the algorithm halts, and returns $t_{11}$ and the derived flow execution scenarios as shown in~(\ref{ex:analysis:3}) for debugger to examine further.

\subsection{Flow Execution Scenarios}
\label{sec:new-scen}

The trace analysis approach in~\cite{Zheng2016ISQED} does not capture orderings among flow instances for execution scenarios.  However, from a debugger's point of view, communication protocols can be related. For example, a firmware loading protocol always happens before a firmware execution protocol. If a firmware execution protocol is found to happen before a firmware loading protocol, that possibly indicates an error in the system implementing such protocols.
Such properties cannot be checked by the previous approach.  

To address that problem, this paper presents a new definition of flow execution scenarios as 
\[ 
\{ (F_{i,j}, s_{i,j}, \mathit{start}_{i,j}, \mathit{end}_{i,j})~|~F_i \in {\bf F}\}
\]
where $\mathit{start}_{i,j}$ and $\mathit{end}_{i,j}$ are two indices representing relative time when $F_{i,j}$ is initiated and completed.  The ordering relations can be derived by comparing their {\em start} and {\em end} indices.  For example, for two flow instances in an execution scenario, $(F_{u,v}, s_{u,v}, \mathit{start}_{u,v}, \mathit{end}_{u,v})$ and 
$(F_{x,y}, s_{x,y}, \mathit{start}_{x,y}$, $\mathit{end}_{x,y})$,  $F_{u,v}$ is initiated before $F_{x,y}$ if $\mathit{start}_{u,v} < \mathit{start}_{x,y}$, or $F_{x,y}$ is initiated after $F_{u,v}$ is completed if $\mathit{end}_{u,v} < \mathit{start}_{x,y}$. The ordering relations can provide more accurate information for understanding system execution under limited observability.  Section~\ref{section:algo} explains how  $\mathit{start}_{i,j}$ and $\mathit{end}_{i,j}$ are decided during the trace analysis.

In order to support the new definition of flow execution scenarios, the trace abstraction, which maps an observed silicon trace to a linear sequence of flow events as in~\cite{Zheng2016ISQED}, is also generalized.  A SoC design can be viewed as a group of IP blocks networked by an on-chip interconnect fabric.  These blocks communicate with each other through communication links, each of which implements a protocol, such as ARM AXI, over a set of wires.  The approach presented in this paper is communication centric in that it works on silicon traces on a selected number of wires of a selected number of communication links for observation.  Suppose that there are $n$ communication links, and some wires from each link are selected for observation.  A silicon trace is assumed to be a sequence of $\alpha_0,\alpha_1,\ldots$ such that each $\alpha_i$ is a vector defined as
\[
\alpha_i = \langle \alpha_{0,i}, \ldots \alpha_{n,i} \rangle
\]
where $\alpha_{k,i}$ is a state on link $k$ in step $i$. 

If all wires of a link are observable, then a state on that link can be uniquely mapped to a flow event of the same link.  Under limited observability, a state on a link is typically mapped to a set of flow events.  Therefore, a silicon trace is abstracted to a sequence $\vec{E_0}, \vec{E_1}, \ldots$ where 
\begin{equation}
\label{eq:trace-abst}
\vec{E_i} = \langle E_{0,i}, \ldots, E_{n, i}\rangle
\end{equation}
is a vector of sets of flow events abstracted from $\alpha_{i}$, and each $E_{k,i}$ in $\vec{E_i}$ is a set of flow events abstracted from state $\alpha_{k,i}$ in $\alpha_i$.  No temporal orderings exist among all events in $\vec{E_i}$.  On the other hand, for two events, $e_i \in \vec{E}_i$ and $e_j \in \vec{E}_j$ such that $i < j$, then $e_i$ happens before $e_j$.  

Based on different levels of information captured, this paper classifies flow execution scenarios as follows.
\vspace{-3pt}
\begin{itemize}\setlength{\itemsep}{0pt}
\item {\em Type-1} execution scenarios capture the number of instances of each flow specification {\em initiated} from a silicon trace, and their relative orderings of initiations. 

\item {\em Type-2} execution scenarios, on top of what is captured by Type-1 scenarios, capture {\em completion} of each flow instance.  This additional information can be used to identify potential problems if there is any flow instance that is not completed.  Furthermore, Type-2 execution scenarios capture the relative orderings among all flow instances as described above.

\item {\em Type-3} execution scenarios, on top of what is captured by Type-2 scenarios,  capture information on execution paths followed by individual flow instances.  This information can provide a means to debuggers to have a detailed examination on {\em how} each flow instance  is executed. 
\end{itemize}
\vspace{-3pt}
These different execution scenarios can be used to provide different views of system execution, from coarse-grained to more detailed ones, at different stages of debug.

\subsection{Algorithms}
\label{section:algo}

Algorithm~\ref{algo:top} shows the top-level procedure for detecting internal flow executions based on a partially observed silicon trace, and checks the compliance \emph{wrt} a given flow specification. It takes as inputs $\mathbf{F}$, a set of system level flow specifications, and a signal trace $\rho$, which is assumed to be a sequence of states on a set of observable trace signals, and each state is uniquely indexed starting from $0$. 

This algorithm  scans trace $\rho$ starting from index~$h$ initialized to $0$, extracts all possible flow events from $\rho$ at index $h$ as described in section~\ref{sec:new-scen}~(line~6), and  maps each of those extracted flow events to update already detected execution scenarios~(line~11). The algorithm terminates if one of two conditions holds.  If an inconsistence is encountered, the set of detected partial execution scenarios along with two indices $h$ and $i$ are returned~(line~16). Index $h$ provides temporal information on when the inconsistency occurs, while $i$ provides spatial information on which communication link an inconsistent event is transmitted.  If no inconsistency is found, the set of all execution scenarios compliant with the observed trace is returned~(line~18) when index $h$ is larger than the length of the trace.  


Algorithm~\ref{algo:flow-analysis} takes the specification $\mathbf{F}$, an execution scenario $\mathit{scen}$, a flow event $e$, and index $h$ of the trace where $e$ is extracted, and it produces a set of execution scenarios $\mathcal{R}$ consistent with $e$.  This algorithm performs two tasks.  In the first task~(lines~5-12), the algorithm checks every flow instance to decide if $e$ can be accepted.  If such an instance is found~(line~7), then it is updated with the new state as the result of $e$~(line~9).  Furthermore, if $e$ causes the flow instance to complete, its index $\mathit{end}_{i,j}$ is set to $h$~(line~10-11), indicating the completion of that instance  due to event $e$ at step $h$ of the trace.  In task~2, all possibilities where $e$ can initiate a new flow instance are considered~(line~14-20).  If a new instance can be initiated, its $\mathit{start}_{i,j}$ is set to $h$, indicating the initiation of that instance due to a signal event at step $h$ of the trace.

\begin{algorithm}[tb]
\setstretch{1.05}
\DontPrintSemicolon
\textcolor{green}{/* $\mathbf{F}$: a set of flow specification */}\\
\textcolor{green}{/* $\rho$: a partially observed silicon trace */}\\

$\mathcal{M} \gets \{\emptyset\}$\;
$h \gets 0$\;
\While{$h \leq |\rho|$} {
	$\vec{E} \gets \mathit{abstract}(\rho, h)$\;
 	\ForEach{$E_i$ of $\vec{E}$} {
        $inconsistent \gets \mathit{true}$\;
 		\ForEach{$e \in E_i$} {
 			\ForEach{$scen \in \mathcal{M}$} {
				$Scens\gets analysis(\textbf{F},\, scen,\, e,\, h)$\;
 				\If{$Scens \neq \emptyset$}{
 					$inconsistent \gets \mathit{false}$\;
 					$\mathcal{M}\gets (\mathcal{M} - scen) \cup Scens $\;
 				}
	 		}
        }
		\If{$inconsistent = \mathit{true}$}{
			\Return $(\mathcal{M}, h, i)$\;
	  	}
    }
    $h \gets h+1$\;
}
\Return $(\mathcal{M}, -1, -1)$\;
\caption{Check-Compliance$(\textbf{F}, \, \rho)$ }
\label{algo:top}
\end{algorithm}

\begin{algorithm}[tb]
\setstretch{1.05}
\DontPrintSemicolon
\textcolor{green}{/* $scen = \{(F_{i,j},\, s_{i,j},\, start_{i,j},\, end_{i,j})\}$ */}\\
\textcolor{green}{/* $e$ is a flow event abstracted from silicon trace at index $h$ */}\\

$\mathcal{R} = \emptyset$\;
\textcolor{green}{/* Check if $e$ can change state any existing flow instances of $scen$ */}\\
\ForEach {$(F_{i,j},\, s_{i,j},\, start_{i,j},\, end_{i,j}) \in scen$} {
	$s'_{i,j} \gets \mathit{accept}(F_{i,j},\, s_{i,j},\, e)$\;
    \If{$s'_{i,j} \not= \emptyset$} {
		Let $scen'$ be a copy of $scen$\;
		Replace $s_{i,j}$ of $scen'$ with $s'_{i,j}$\;
    	\If{$s'_{i,j} = F_i.s_{end}$} {
			Update $end_{i,j}$ of $scen'$ with $h$\;
		}        
		$\mathcal{R}\gets \mathcal{R}  \cup scen'$\;
	}
}

\textcolor{green}{/* Check if $e$ can extend $scen$ by initiating new flow instances */}\\
\ForEach {$F_i \in \mathbf{F}$} {
	create a new instance $F_{i, h}$ \;
    $s'_{i,h} \gets \mathit{accept}(F_{i,h},\; F_i.s_0,\; e)$\;
    \If{$s'_{i,h} \not= \emptyset$} {
   		Let $scen'$ be a copy of $scen$\;
		$scen' \gets scen' \cup (F_{i,h},\; s'_{i,h},\; h,\; -1)$ \;
		$\mathcal{R}\gets \mathcal{R}  \cup scen'$\;
	}
}

\Return $\mathcal{R}$
	
\caption{$Analysis(\textbf{F},\, scen,\, e,\, h)$ }
\label{algo:flow-analysis}
\end{algorithm}

\subsection{On the Complexity and Accuracy}
\label{sec:complexity}


Due to the limited observability, reconstructing system level executions from an observed silicon trace is an imprecise process.  The large number of execution scenarios typically derived during the analysis would take large amounts of runtime and memory to process and to store, thus making it less efficient.  This is referred to as the {\em complexity} problem of the trace analysis.  After the analysis is done, a large number of derived execution scenarios make it difficult to understand the analysis results, thus being less helpful for debugging.  Obviously, a single flow execution scenario derived at the end of the trace analysis provides much more precise information for debug than ten candidate flow execution scenarios.  This is referred to as the \emph{accuracy} problem of the trace analysis.  

The contributing factors to the complexity and accuracy problems are explained below. 
\vspace{-3pt}
\begin{enumerate}\setlength{\itemsep}{0pt}
\item {\em A signal event mapped to a set of flow events} $-$
Due to the limited observability, a signal event of an observed silicon trace is often interpreted as a number of different flow events, which typically leads to derivation of a number of different execution scenarios.  This situation is exacerbated by the fact that silicon traces are often very long, which could lead to excessively large numbers of possible execution scenarios derived during or at the end of the analysis.  

\item {\em A flow event mapped to different temporal flow instances} $-$ 
Temporal flow instances refer to the flow instances activated by the same component,  {\em e.g.} read/write flows activated by {\tt CPU\_0}.  If several temporal instances of some flows are activated by a component, mapping flow events to those flow instances can be ambiguous.   For example, suppose that an execution scenario includes two instances of the flow as shown in Figure~\ref{fig:flow-spec-ex} activated by {\tt CPU\_0}, one in state $\{p_2\}$, and the other one in state $\{p_8\}$.  An instance of flow event ${\tt (Cache\_0:CPU\_0:wr\_resp)}$ can be mapped to either flow instance leading to two new execution scenarios from the current one.

\item {\em A flow event mapped to flow instances activated by different components} $-$ This situation can happen when flow instances that share some common events are activated by different components.    For example, suppose an execution scenario has two instances of the flow as shown in Figure~\ref{fig:flow-spec-ex}, one activated by {\tt CPU\_0} and the other one by {\tt CPU\_1}, and both are in state $\{p_6\}$.  A flow event ${\tt (Mem:Bus:rd\_resp)}$ can be mapped to either one of these two instances, leading to two new execution scenarios derived from the current one. 
\end{enumerate}
\vspace{-3pt}

The above issues can be mitigated by good signal selections to be discussed in the following section.  In order to evaluate the impacts of different trace signal selections on the complexity and accuracy of the trace analysis, this paper introduces two quantitative metrics.  The complexity is measured by the \emph{peak} count of flow execution scenarios encountered during the analysis process, {\em i.e.}, the largest size of $\mathcal{M}$ encountered during the execution of Algorithm~\ref{algo:top}.  The  accuracy is measured by the \emph{final} count of flow execution scenarios derived at the end of the analysis process, {\em i.e.}, the size of $\mathcal{M}$ returned on either line~16 or 18 of Algorithm~\ref{algo:top}.

\section{Trace Signal Selection}
\label{sec:sigselect}

Trace signal selection is a critical step in post-silicon debug.  It includes two different efforts: pre-silicon and post-silicon.  During pre-silicon selection, a few thousand signals among a vast number of internal signals are tapped for observation.  All necessary signals must be selected at this stage, otherwise, expensive re-design along with silicon re-spin are required.  During post-silicon debug, a small subset of those tapped signals are routed to the chip interface for tracing during system execution.

Previous work such as~\cite{mishra2011vlsi} is typically applied to gate level design models, and the quality of the results is evaluated by the commonly used state restoration ratio.  However, it is difficult to scale those methods to large and complex SoC designs. More importantly, signals selected at the gate level are often irrelevant to system-level functionalities.  There is an attempt to raise the abstraction level for trace signal selection to the register transfer level~(RTL) guided by assertions~\cite{Ma2015ICCAD}, however that work does not consider system level functionalities either.  
In~\cite{amrein2015system}, a system level protocol guided approach is proposed.  It is similar to our work in that both are based on system level protocols.  However, the selection techniques developed in~\cite{amrein2015system} are simple and irrelevant to understanding silicon traces at the system level, and the evaluation was performed on an abstract transaction level model.

\begin{figure}[tb]
\begin{center}
\resizebox{2.2in}{!}{
\includegraphics[width=.4\textwidth]{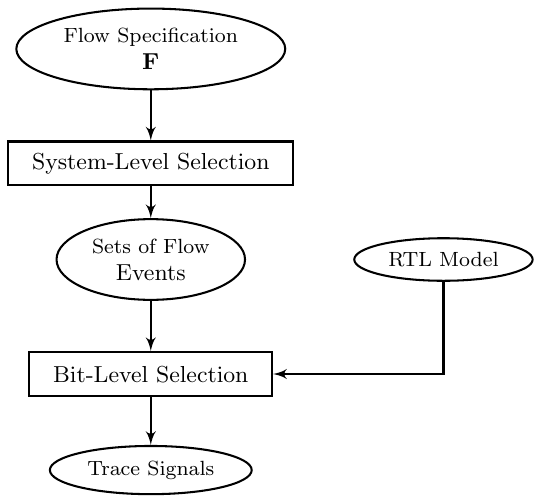}
}
\end{center}
\caption{A framework of trace signal selection.}
\label{fig:sigselec}
\end{figure}

This section introduces a framework shown in~Figure \ref{fig:sigselec} for trace signal selection guided by system-level protocols. Due to the page limit, this paper only considers the pre-silicon trace signal selection.  Since the pre-silicon selection needs to support all types of execution scenarios,  it is sufficient to consider only Type-3 scenarios as they supersede Type-1 or -2 scenarios.

\begin{figure}[tb]
\begin{center}
\begin{tabular}{cc}
\resizebox{1in}{!}{
\includegraphics[width=.2\textwidth]{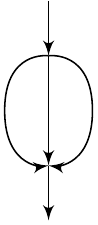}
}

&

\resizebox{.5in}{!}{
\includegraphics[width=.2\textwidth]{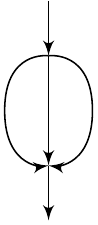}
}
\\
(a) & (b)
\end{tabular}
\end{center}
\caption{Examples of flow structures.}
\label{fig:flow-topo}
\end{figure}

\subsection{System Level Selection}
\label{sec:sys-selec}

During the system level selection, different subsets of flow events for observation are selected from given flow specifications.  Then, those results are passed to the more refined bit level selection.

To support Type-3 scenarios, the start and end events of all flow specifications must be selected.  If a flow specification has multiple branches, additional events may need to be selected so that the branch followed by a flow instance during system execution can be captured.  Figure~\ref{fig:flow-topo} shows two examples of different branching structures for flows.
\begin{itemize}\setlength{\itemsep}{0pt}
	\item In Figure~\ref{fig:flow-topo}(a), each branch ends with an unique event.  There is no need to select additional events as observing different end events can clearly identify the branch followed during system execution.
	\item In Figure~\ref{fig:flow-topo}(b), branches split and then join, and the flow ends with a common event.  In this case, an unique event needs to be selected for each branch.  
\end{itemize}

The flow shown in~Figure~\ref{fig:flow-spec-ex} has three branches with a structure similar to Figure~\ref{fig:flow-topo}(b).  Its start and end events are $\{t_1, t_8, t_9, t_{10}\}$.  Note that $t_8$, $t_9$, and $t_{10}$ actually refer to the same event.  There is no choice for the right branch as $t_{10}$ must be selected.  To identify the left two branches from the right one, either $t_2$ or $t_3$ needs to be selected.  Similarly, one of events in $\{t_4, t_5, t_6, t_7\}$ needs to be selected in order to identify the left branch from the middle one.  Therefore,  all possible event selections for that flow are 
\[
\{t_1, t_8, t_9, t_{10}\} \times \{t_2, t_3\} \times \{t_4, t_5, t_6, t_7\}.
\]

Among possibly large number of event selections, there are two types of events that have interesting characteristics.  One type includes events that are unique to specific flows~(ref. unique events), while the other type includes events shared by multiple flows~(ref. shared events).   This section considers their impacts on the complexity and accuracy of the trace analysis and the signal selection.  For the complexity and accuracy of the trace analysis, only issues \#2 and \#3 in section~\ref{sec:complexity} are considered in this section. Issue~\#1 is relevant to the bit level selection. 

It is important to select events that can be mapped to smallest number of flow instances during the trace analysis in order to reduce the complexity and improve the accuracy.  Refer to the flow shown in Figure~\ref{fig:flow-spec-ex}.  Events $t_2$ and $t_3$ are used only in that flow.  During the trace analysis, they are just mapped to the instances of that particular flow.  Issue~\#2 can be addressed in that instances of different flows initiated by the same component are ignored for those events.   Issue~\#3 is addressed in that instances of any flows initiated by different components are ignored for them.  $t_4$ and $t_7$ have a similar characteristic. 

On the other hand, events $t_5$ and $t_6$ are used in many different flows of different components.  Those flows can be read/write flows of {\tt CPU\_0} or {\tt CPU\_1}.  During the trace analysis, if there are multiple instances of such flows, it is impossible to know which of those flow instances cause those events to be generated.  Therefore, the analysis algorithm has to map those events to those flow instances in all possible ways.   That can cause huge negative impacts on the complexity and accuracy of the trace analysis. 

In terms of trace signal selection,  those two types of events can lead to different results.  If unique events are selected.  then the total number of events selected can be large, and as a result, a large number of trace signals need to be selected in order to observe those events.  On the other hand, the total number of events can be smaller if shared events are selected.  That leads to a  smaller number of trace signals that need to be selected.  The negative impacts of selecting shared events can also be mitigated if certain implementation details are available.  Next section gives more discussions on that point.

\subsection{Bit Level Selection}
\label{sec:sig-sel}

The bit level selection takes as inputs the set of event selections produced in the previous step and an RTL model that implements the system flow specifications, and performs two tasks for each event selection:
\vspace{-3pt}
\begin{enumerate}\setlength{\itemsep}{-1pt}
\item Evaluate its quality {\em wrt} the three issues discussed in Section~\ref{sec:complexity};
\item Choose one selection, and generate a set of candidate trace signals that implement the selected events.
\end{enumerate}
\vspace{-3pt}
The ultimate goal of the bit level selection is to produce a reduced set candidate trace signals optimized for the trace analysis approach.  Since the bit level selection depends on implementation specifics, this section can only discuss some general guidelines and tradeoffs.  Note that flow specifications are typically independent of memory address and data information.  Therefore, the address and data bits included in event implementations can be generally ignored.


Signals that implement the {\tt Cmd} field of flow events are selected based on their respective \emph{distinguishing power}.  Given a set of flow events $E$ and a set of signals $W$ that implement $E$, the distinguishing power of $W_i \subseteq W$, is defined by $E$ can be partitioned {\em wrt} $W_i$.  A finer partition means higher distinguishing power.  For example, suppose two flow events on link $({\tt cpu0, DCache0})$ implemented by eight signals $b_7\ldots b_0$ with the following encodings.
\[
\begin{array}{ll}
({\tt cpu0:DCache0:wr\_req)} & 0100\;0000\\
({\tt cpu0:DCache0:rd\_req)} & 1000\;0000
\end{array}
\]
Under these encodings, signals $b5\ldots b_0$ have zero distinguishing power. $b_7$ and $b_6$ have the equal power, therefore selecting either one would be fine.  Selecting signals with high distinguishing power helps to address issue~\#1 as discussed in Section~\ref{sec:complexity}.

RTL models may contain additional implementation information that can help to address issue \#2 and \#3.  For example, memory operations may be executed out-of-order.  In this case, CPUs usually assign unique sequence IDs to flow instances to maintain data and control dependency in the original programs. If sequence IDs are available, selecting signals implementing them can help address issue \#2.  

If the on-chip interconnect needs to handle events from different components in a system, the events are usually assigned with tags to identify their originating components.  Selecting tags can affect how events are selected.  Refer to Figure~\ref{fig:flow-spec-ex} for the following discussion.
\vspace{-3pt}
\begin{enumerate}\setlength{\itemsep}{1pt}
	\item If unique events such as $t_4$ or $t_7$ are selected, observing tags is not needed.  

	\item Shared events $t_5$ or $t_6$ are selected along with tags.
\end{enumerate}
\vspace{-3pt}
For option~2, tags can help to map events to the flow instances with the same tags during the trace analysis, thus addressing issue \#3.  Even though additional signals for tags are selected, the total number of events may be smaller if the shared events are used in many different flows, therefore resulting in reduced signals for observation overall. 

The following discussion illustrates yet another example of how implementation information can allow different events to be selected.  Refer to Figure~\ref{fig:flow-spec-ex}.  That flow contains two branching places, $p_2$ and $p_4$.  When a flow instance reaches $p_2$, which branch to take next depends on whether the cache operation is hit or miss.  Similarly, which branch to take at $p_4$ depends on whether the cache snoop operation is hit or miss.  If these two status signals are available and included for observation, there is no need to select branch events.  Observing start/end events plus those status signals are sufficient to identify branches followed by a flow instance during system execution.

\section{Experimental Results}
\label{section:results}

To the best of our knowledge, this work is the first to present a systematic approach to post-silicon trace analysis guided by system level protocols.  We are not able to find any similar previous work where ours can be evaluated and compared with.  The closest work to ours is \cite{Zheng2016ISQED}.  However, our work is more general and developed with practical considerations.  Additionally, the work in~\cite{Zheng2016ISQED} is discussed and evaluated based on an abstract transaction-level model while our approach is evaluated on a RTL model.


\subsection{The Model}


The ideas and techniques presented in this paper are evaluated on a multi-core SoC prototype, as shown in Figure~\ref{rtlstruc}, which implements a number of common industrial system-level protocols including cache coherence and power management. This prototype is a cycle- and pin-accurate RTL model written in VHDL.  Even though this model is simple compared to real SoC designs, it is much more sophisticated than the gate-level benchmark suites typically considered as targets for post-silicon analysis~\cite{ko,mishra2011vlsi,valeria}.

Since the proposed trace analysis approach is communication centric, the focus of this model is the implementation of system-level protocols.  The CPUs are treated as a test environment where software programs are simulated in VHDL to trigger various protocols.  Therefore, there is no instruction cache as no instructions are involved when the CPUs are simulated.  The peripheral blocks, GFX, PMU, Audio, \emph{etc}, are also described as abstract models that generate events to initiate flows or to respond incoming requests. 

More details of some system-level protocols implemented in our model can be found in~\cite{amrein2015system}.  They include downstream read/write protocols for each CPU, upstream read/write for the peripheral blocks, and system power management protocols, which are abstracted from real industrial protocols.  These system-level protocols are supported by inter-block communication protocols based on the ARM {AXI4-lite~\cite{AXI4}}.  A total of {$16$} flows are implemented for this prototype.







A flow event is generated from a source and consumed by a destination by messages transmitted over that link.  In our model, each message is organized as follows.
\[
\langle {\tt Val(1),\ Cmd(8),\  Tag(8),\ Sid(8),\ Addr(32),\ Data(32)} \rangle
\]
The meanings of the message fields are given below.  The numbers following the individual fields indicate their respective widths.  Note that not all fields are used on all links. That model has over four thousand single bit signals.   
\begin{itemize}\setlength{\itemsep}{-1pt}
\item[{\tt Val}] indicates validity of a message. 
\item[{\tt Cmd}] carries operations to be performed by the target block. 
\item[{\tt Tag}] is used by {\tt Bus} to identify the original sources of messages from different blocks that go to the same destination, \emph{e.g.} memory {\tt wr\_req} from {\tt Bus} in response to {\tt wr\_req} from both CPUs. 
\item[{\tt Sid}] is an unique number generated by a component to represent sequencing information of flows initiated by the same component. 
\item[{\tt Addr}] carries the memory address at the target block where {\tt Cmd} is applied. 
\item[{\tt Data}] carries data to a target or from a source.  Its width can vary depending on the links where a message is sent. On the links between {\tt Cache} to {\tt Bus}, the width is equal to the size of the cache block, which is $64$ bytes. For all the other links, the width is $32$ bits.
\end{itemize}

\subsection{Experiment Setup}

\noindent{\bf Test Environment}~~The prototype is simulated in a random test environment where CPUs, GFX, and other peripheral blocks are programmed to randomly select a flow to initiate in each clock cycle.  The contents of {\tt Cmd}, {\tt Addr}, and {\tt Data} in each activated flow are set randomly.  Additionally, CPUs can activate power management protocols non-deterministically.  Each of these blocks activates a total of {100} flow instances during entire simulation.

\vspace{3pt}

\noindent{\bf Trace Signal Selection}~~
In the experiments, different selections of trace signals are produced as discussed in~section~\ref{sec:sigselect}, and their impacts on the complexity and accuracy of the trace analysis approach are evaluated.   The list below explains the selections at the system level while information on the bit level selection is given in~Table~\ref{tab:test1-result}.

\vspace{-2mm}

\begin{itemize}\setlength{\itemsep}{-3pt}
 \item[S1] All events of all flow specifications, and all signals implementing each event are selected. This selection offers full observation, and provides a baseline for comparing with other selections. 

 \item[S2] The start and end events of all protocols are selected.  Furthermore, for each branch in each flow, one unique event is selected. 



\item[S3] The start and end events of all protocols are selected.  Furthermore, for each branch in each flow, a highly shared event is selected. 
 
\item[S4] The start and end events of all protocols are selected. Instead of selecting events for branches in each flow, signals whose states control the flow branching are selected.

\end{itemize}

\vspace{-2mm}

At the bit level, the {\tt Addr} and {\tt Data} fields are not considered. On the other hand, the {\tt Val} bit is always selected so that valid messages can be identified from observed traces.  For selections S2, S3, and S4, experiments are performed to evaluate all combinations of {\tt Cmd}, {\tt Tag} and {\tt Sid} fields. 


\subsection{Result Analysis}

\begin{table*}[tb]
\caption{Runtime Results of Trace analysis with different trace signal selections. Runtime is in seconds and memory usage is in MB. $-$ indicates the results are not available due to the 10 minute time limit exceeded.}

\begin{center}
\resizebox{7in}{!}{
\begin{tabular}[c]{|m{1.4cm}|c|c|c|c|c|c|c|c|c|c|c|c|c|c|c|c|c|}
\hline

System level selection	&\multicolumn{1}{|c|}{S1} & \multicolumn{5}{|c|}{S2}&\multicolumn{5}{|c|}{S3}&\multicolumn{5}{|c|}{S4}  \\
\hline


  & & & & & & U\; S & & & & & U\; S & & & & &  U\; S\\
 {\tt Cmd} &\checkmark     & \checkmark &&\checkmark&\checkmark	& \checkmark  \  \ \ \ \    &\checkmark	&&\checkmark&\checkmark & \checkmark \   \ \ \ \      &\checkmark&&\checkmark&\checkmark&\checkmark \  \ \ \ \   \\
{\tt Tag} & \checkmark & \checkmark & \checkmark & & \checkmark & \ \  \ \  \checkmark &\checkmark &\checkmark&&\checkmark&\ \  \ \  \checkmark & \checkmark &\checkmark&&\checkmark&\ \   \ \ \checkmark \\
{\tt Sid} & \checkmark & \checkmark & \checkmark & \checkmark &  & \checkmark\ \checkmark & \checkmark & \checkmark &\checkmark& &\checkmark\ \checkmark &\checkmark &\checkmark&\checkmark&&\checkmark\ \checkmark \\ 
 \hline
 \hline
\# Bits & 870 & 545 & 401 & 401 & 401 & 401 & 495 & 367 & 367 & 367 & 367 & 378 & 258 & 258 & 258 & 258\\
\hline
\hline
\# scen & \multirow{2}{*}{1}&\multirow{2}{*}{1} &  \multirow{2}{*}{$-$}& \multirow{2}{*}{1}&\multirow{2}{*}{$-$} &\multirow{2}{*}{1} & \multirow{2}{*}{1} & \multirow{2}{*}{$-$}& \multirow{2}{*}{1}&\multirow{2}{*}{$-$} &\multirow{2}{*}{1} & \multirow{2}{*}{1} &\multirow{2}{*}{$-$} & \multirow{2}{*}{1}&\multirow{2}{*}{$-$}&\multirow{2}{*}{1}  \\

(Final) & & & &&&& &&&& &&&&&\\
\hline
\# scen& \multirow{2}{*}{1}&\multirow{2}{*}{1} & \multirow{2}{*}{$>1M$}& \multirow{2}{*}{5184}& \multirow{2}{*}{$110k$}&\multirow{2}{*}{1} & \multirow{2}{*}{1} &\multirow{2}{*}{$>4M$} & \multirow{2}{*}{5184}&\multirow{2}{*}{$221K$} &\multirow{2}{*}{1} &\multirow{2}{*}{1} &\multirow{2}{*}{$>1M$} & \multirow{2}{*}{8}&\multirow{2}{*}{$>8M$}&\multirow{2}{*}{1}  \\
(Max)  & & & &&&& &&&& &&&&&\\
\hline
Time & 1.628 & 1.475 & 600 & 3.679 & 600 & 1.464 &     1.444 & 600 & 3.812 & 600 & 1.426   & 1.430 & 1.411 & 1.424 & 600 & 1.419 \\
\hline
Mem  & 0.516 & 1.10 & $>2 GB$ & 4.2 & 66 & 1.124 &  1.11 & $>5 GB$ & 4.2 & 101 & 1.1 & 0.504  & $>2$ GB & 0.58& $>5$ GB & 1.116 \\
\hline
\end{tabular}
}
\end{center}
\label{tab:test1-result}
\end{table*}%

In Table \ref{tab:test1-result}, a \checkmark means that all signals implementing a particular field for all selected events in selection $S_X$ are traced.  Otherwise, all those signals are not traced.  Third row (\# Bits) shows the total numbers of single-bit signals are traced for different selections. As discussed in section~\ref{sec:sigselect}, system-level selection may choose events unique to particular flows or events shared by multiple flows.  From the table, we can see that selecting shared events leads to a smaller number of trace signals (S3) compared with selecting unique events (S2).  However, if status signals controlling flow branching are selected without selecting any branch events, S4 leads to the smallest trace signal selection.

From the table, it is quite obvious that not selecting {\tt Cmd} or {\tt Sid} has severe impacts on the trace analysis as explained in issues \#1 and \#2 in section~\ref{sec:complexity}.  On the other hand, not selecting {\tt Tag} has negative impacts, but not as severe.  The trace analysis can still finish even though it takes more time and memory. Next, compare the results obtained by selecting {\tt Cmd} and {\tt Sid} but no {\tt Tag} under S2$-$S4. The results with S4 are much better than S2 or S3.  This is due to that no branch events are selected for S4, therefore, issues \#2 and \#3 are avoided.  Combined with the benefit of reduced trace signals, S4 appears to be the best option.  On the other hand, not selecting any branch events may cause difficulty in understanding flow execution if a branch is long and a system execution fails to reach the end of that branch.      

In the above discussion, selections of {\tt Cmd}, {\tt Tag} and {\tt Sid} are applied to all events as the result of the system-level selection.  A finer selection can be used to reduce trace signals if unique events and shared events are considered separately.  For unique events, the sources where they are generated are known from flows, therefore {\tt Tag}s need not be traced.  Shared events may be results of flow instances initiated by different components, therefore tracing {\tt Tag}s are necessary.  On the other hand, tracing or not tracing {\tt Cmd}s has little impact on the trace analysis.  These points are supported by the results shown in columns under $``\mbox{U}\; \mbox{S}''$.  Under S2, compare the results under $``\mbox{U}\; \mbox{S}''$ against those with all three fields selected.  We can see that the runtime performance and the complexity and accuracy of the trace analysis are similar while the trace signals are reduced with the finer selection.  Comparing the results under $``\mbox{U}\; \mbox{S}''$ against those obtained with only {\tt Cmd} and {\tt Sid} selected, the complexity is significantly dropped.  The same conclusion can be drawn for S3 and S4. 

From the above discussion, it is necessary to trace signals implementing {\tt Cmd} and {\tt Sid} whenever possible, and trace as many signals implementing {\tt Tag} as allowed to reduce complexity of the trace analysis even more.  If {\tt Tag} or {\tt Sid} is not part of the design, we recommend to add DFx circuitry in order to trace such information.  In the above experiments, the final execution scenarios under different signal selection, if available, contain the correct number of flow instances initiated, and the orderings among the flow instances, as generated by the test environment, are correctly captured.

\section{Related Work}

Our work is closely related to communication-centric and
transaction based debug.  An early pioneering work is
described by Goossens {\em et al.}~\cite{Goossens2007NOCS,Vermeulen2009VLSI-DAT,Goossens2009DATE}, which advocates the
focus on observing activities on the interconnect network
among IP blocks, and mapping these activities to
transactions for better correlation between computations and
communications.  
A similar transaction-based debug approach is presented
by Gharebhagi and Fujita~\cite{Gharehbaghi2012ISQED}.  It proposes an
automated extraction of state machines at transaction level
from high level design models. From an observed failure trace, it tries to derive a set of feasible transaction traces that lead to the observed failure state.  However, this approach requires manual inputs and may not be able to derive such traces.

Singerman {\em et al.}~\cite{Singerman2011DAC} deploys a central repository of system events and simple transactions defined by architects and IP designers.
It spans across a wide spectrum of the post-silicon
validation including DFx instrumentation, test generation,
coverage, and debug.  
Also, Abarbanel {\em et al.}~\cite{Abarbanel2014DAC} propose a model at a higher-level of abstraction, {\em flows},
is proposed. Flows are used to specify more sophisticated
cross-IP transactions such as power management, security,
etc, and to facilitate reuse of the efforts of the
architectural analysis to check HW/SW implementations.

\section{Conclusion}

An improved trace analysis approach for post-silicon debug is presented where observed raw silicon traces are interpreted {\em wrt} system flow specifications.  In this approach, a new formulation of flow execution scenarios is described where more diverse information among flows can be captured and represented. 
A trace signal selection framework is also described in support of the proposed trace analysis approach.  Some observations on trace signal selections and their impacts on the accuracy and efficiency of the trace analysis are discussed.  Experiments on a non-trivial SoC prototype reveal insights on impacts of different signal selections on the complexity and accuracy of the trace analysis. 
In the future, we plan to perform more extensive and in-depth study on trace signal selections guided by system flow specifications.




%

\begin{thebibliography}{10}

\bibitem{AXI4}
Amba axi and ace protocol specification.
\newblock http://www.arm.com.

\bibitem{Abarbanel2014DAC}
Y.~Abarbanel, E.~Singerman, and M.~Y. Vardi.
\newblock Validation of soc firmware-hardware flows: Challenges and solution
  directions.
\newblock In {\em Proceedings of DAC'14}, pages 2:1--2:4, 2014.

\bibitem{amrein2015system}
M.~Amrein.
\newblock System-level trace signal selection for post-silicon debug using
  linear programming.
\newblock Master's thesis, Univ. of Illinois Urbana-Champaign, May 2015.

\bibitem{mishra2011vlsi}
K.~Basu and P.~Mishra.
\newblock Efficient trace signal selection for post silicon validation and
  debug.
\newblock In {\em VLSI Design (VLSI Design)}, pages 352--357. IEEE, 2011.

\bibitem{valeria}
D.~Chatterjee, C.~McCarter, and V.~Bertacco.
\newblock Simulation-based signal selection for state restoration in silicon
  debug.
\newblock In {\em {\small ICCAD}}, pages 595--601. \small IEEE, 2011.

\bibitem{Foster2015DAC}
H.~D. Foster.
\newblock Trends in functional verification: A 2014 industry study.
\newblock In {\em DAC}, pages 48:1--48:6, 2015.

\bibitem{Gharehbaghi2012ISQED}
A.~M. Gharehbaghi and M.~Fujita.
\newblock Transaction-based post-silicon debug of many-core system-on-chips.
\newblock In {\em ISQED}, pages 702--708, 2012.

\bibitem{Goossens2009DATE}
K.~Goossens, B.~Vermeulen, and A.~B. Nejad.
\newblock A high-level debug environment for communication-centric debug.
\newblock In {\em Proceedings of DATE'09}, pages 202--207, 2009.

\bibitem{Goossens2007NOCS}
K.~Goossens, B.~Vermeulen, R.~v. Steeden, and M.~Bennebroek.
\newblock Transaction-based communication-centric debug.
\newblock In {\em Proceedings of NOCS'07}, pages 95--106, 2007.

\bibitem{ko}
H.~F. Ko and N.~Nicolici.
\newblock Algorithms for state restoration and trace-signal selection for data
  acquisition in silicon debug.
\newblock {\em IEEE TCAD}, 28(2):285--297, 2009.

\bibitem{Ma2015ICCAD}
S.~Ma, D.~Pal, R.~Jiang, S.~Ray, and S.~Vasudevan.
\newblock Can't see the forest for the trees: State restoration's limitations
  in post-silicon trace signal selection.
\newblock ICCAD '15, pages 1--8, Piscataway, NJ, USA, 2015. IEEE Press.

\bibitem{Patra2007}
P.~Patra.
\newblock On the cusp of a validation wall.
\newblock {\em IEEE Des. Test}, 24(2):193--196, Mar. 2007.

\bibitem{Singerman2011DAC}
E.~Singerman, Y.~Abarbanel, and S.~Baartmans.
\newblock Transaction based pre-to-post silicon validation.
\newblock In {\em Proceedings of DAC'11}, pages 564--568, 2011.

\bibitem{Vermeulen2009VLSI-DAT}
B.~Vermeulen and K.~Goossens.
\newblock A noc monitoring infrastructure for communication-centric debug of
  embedded multi-processor socs.
\newblock In {\em VLSI-DAT '09}, pages 183--186, 2009.

\bibitem{Zheng2016ISQED}
H.~Zheng, Y.~Cao, S.~Ray, and J.~Yang.
\newblock Protocol-guided analysis of post-silicon traces under limited
  observability.
\newblock In {\em Proceedings of ISQED'16}, pages 301--306, March 2016.

\end{thebibliography}
%

%

\clearpage

\appendix
\section{CPU read/write downstream protocol}
\begin{figure}[h]
{\color{red}X}=\{ 0, 1\}

{\color{red}X'}=1-X

{\color{red}Target}=\{ Memory, USB, UART, AUDIO, GFX\}

{\color{red}CMD}\ =\{read, write \}
\begin{center}
\resizebox{0.5\textwidth}{!}{
\includegraphics[width=.5\textwidth]{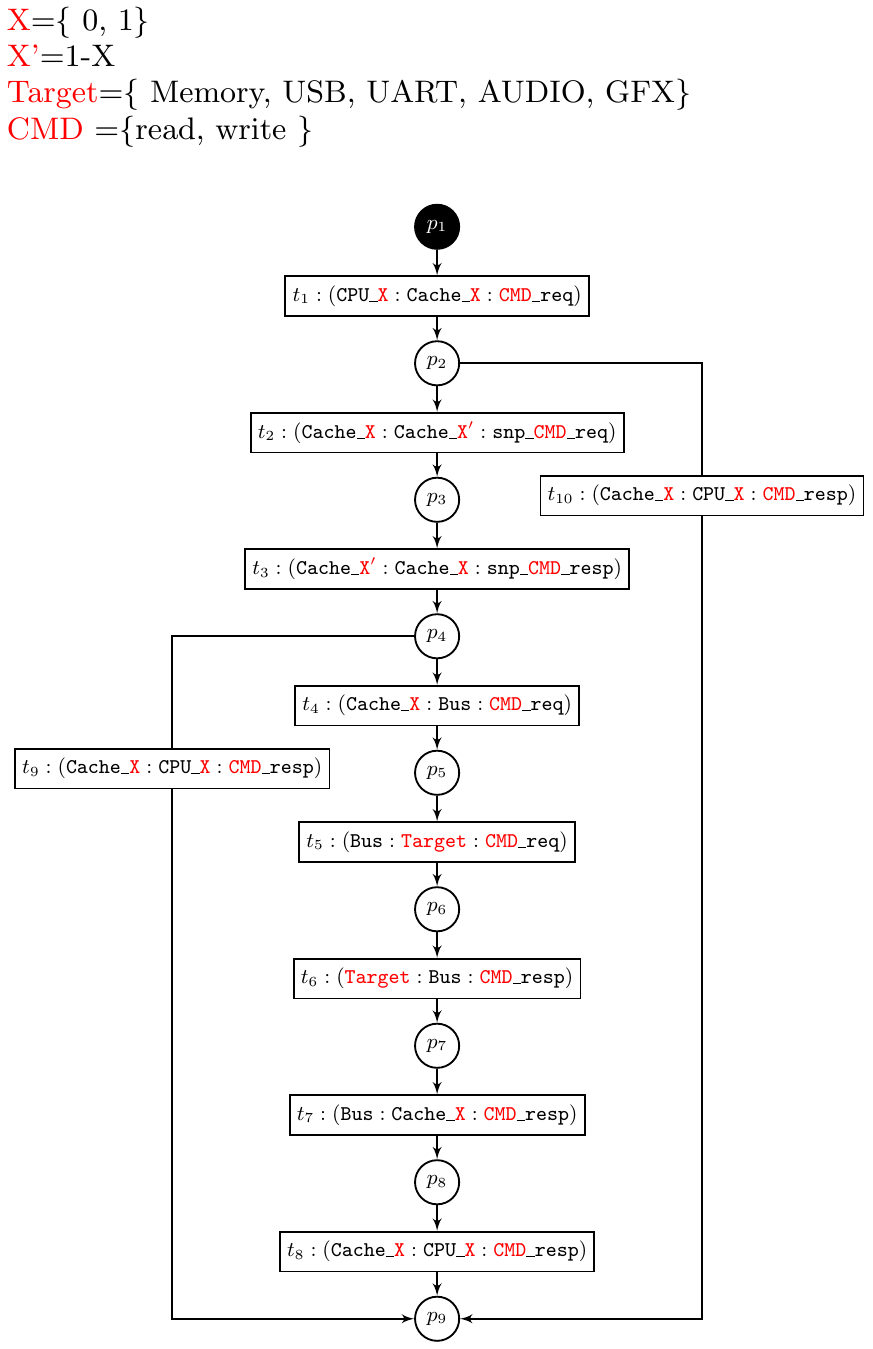}
}
\caption{LPN formalization of a CPU read/write protocol.
}
\label{flow-spec-ex1}
\end{center}
\end{figure}

\begin{figure}[h]
\begin{center}
\includegraphics[width=.5\textwidth]{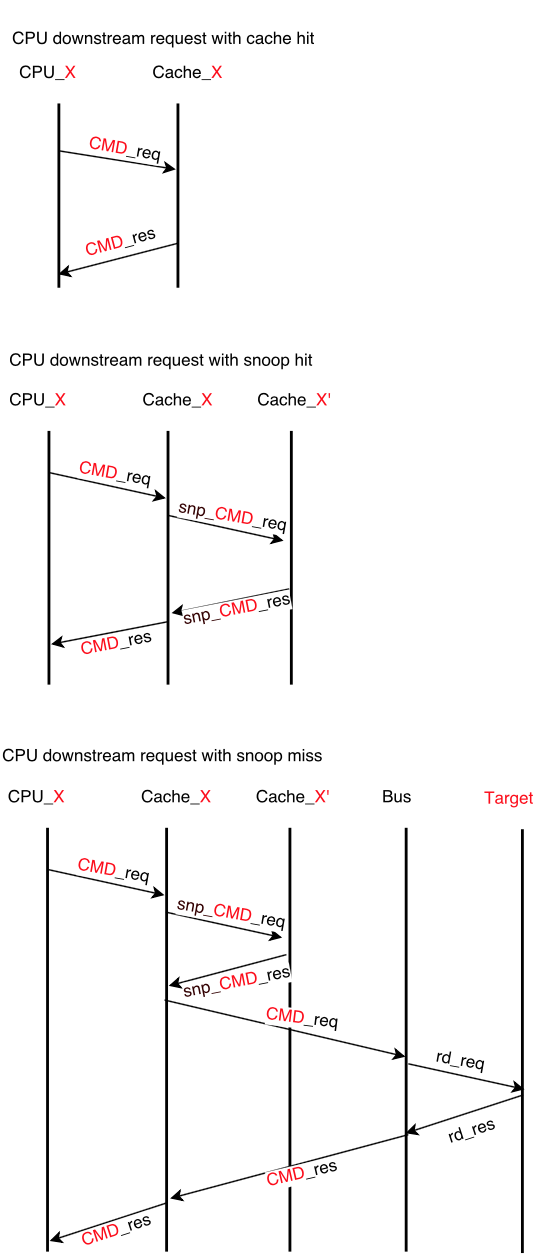}
\end{center}
\caption{MSQ of CPU downstream read/write protocol}
\end{figure}

\clearpage

\section{Upstream read/write protocol}
\begin{figure}[h]
{\color{red}Initiator}=\{ GFX, USB, AUDIO, UART\}

{\color{red}Target}=\{ Memory, USB, UART, AUDIO, GFX\}

Note that a peripheral can't initialize a read flow to read itself

\begin{center}
\resizebox{0.5\textwidth}{!}{
\includegraphics[width=.5\textwidth]{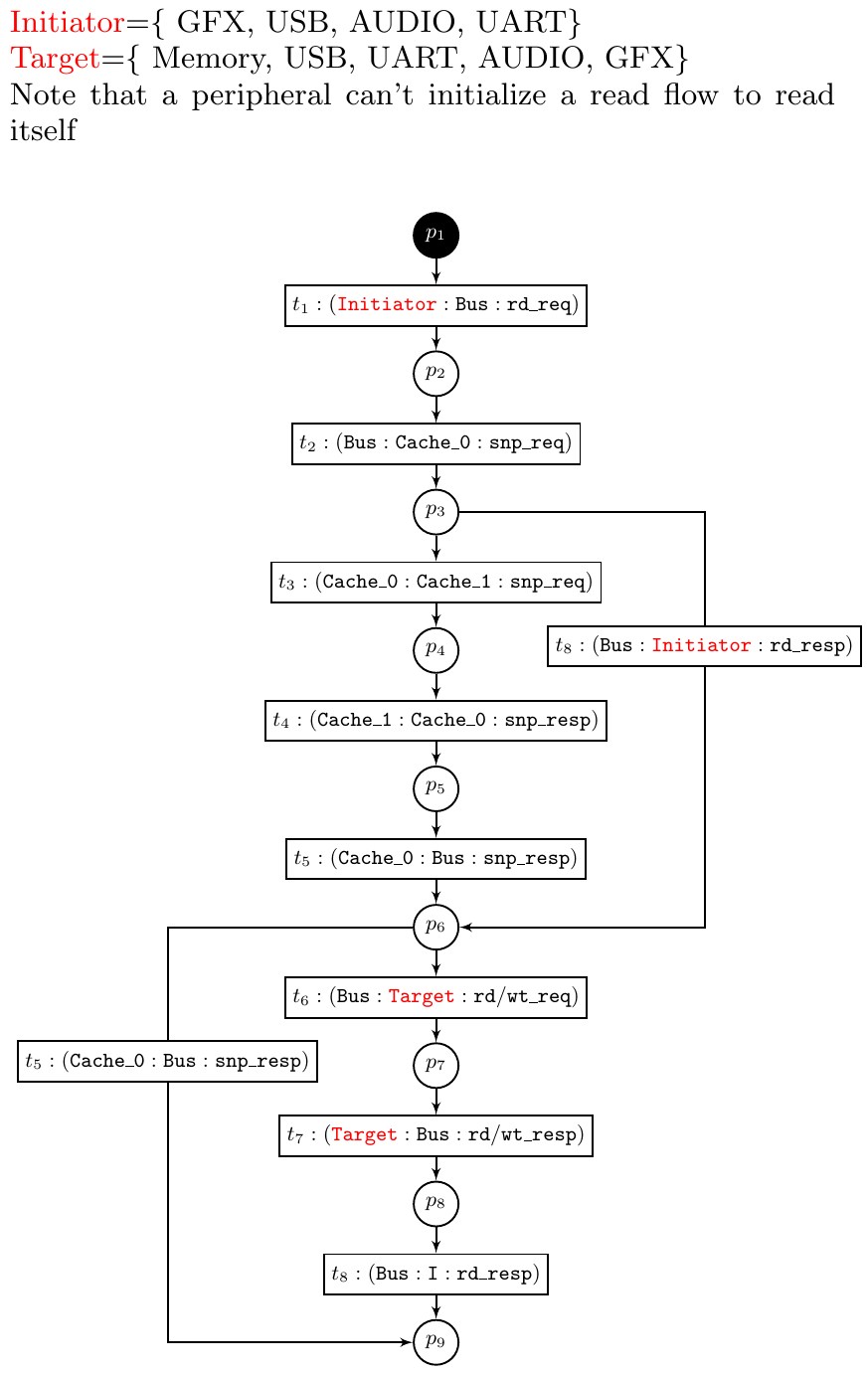}
}
\caption{LPN formalization of a upstream read protocol.
}
\label{flow-spec-uprd}
\end{center}
\end{figure}

\begin{figure}[h]
\begin{center}
\includegraphics[width=.5\textwidth]{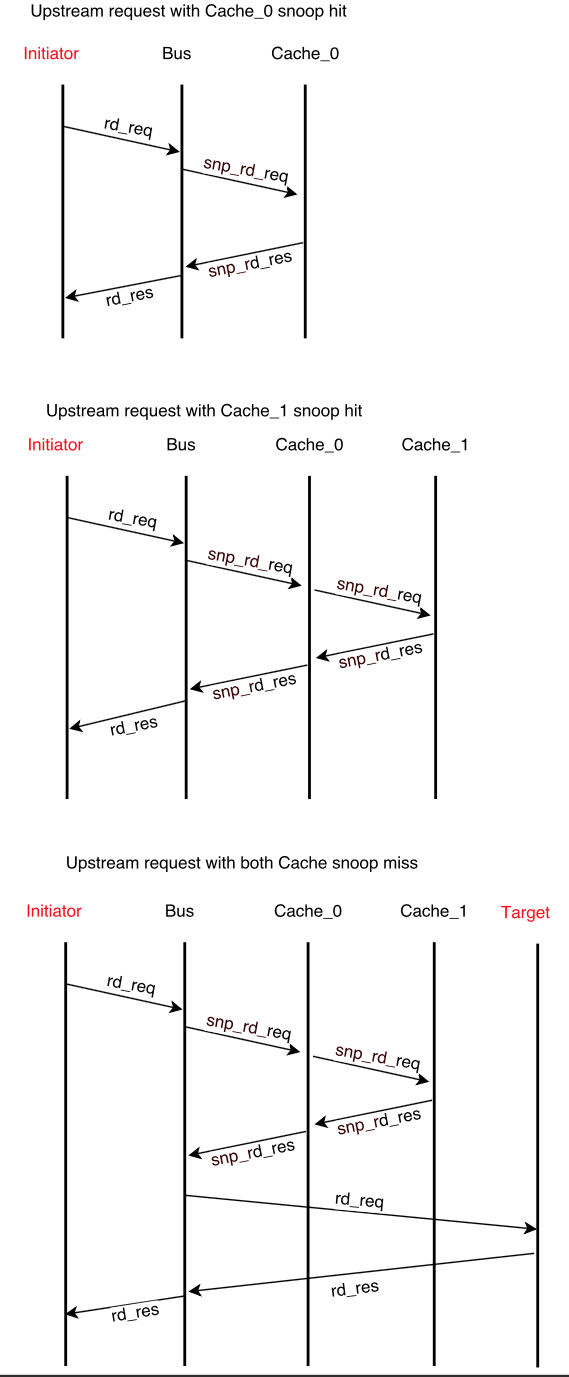}
\end{center}
\caption{MSQ of upstream read protocol}
\end{figure}%

\begin{figure}[h]
{\color{red}Initiator}=\{ GFX, AUDIO\}

{\color{red}Target}=\{ Memory, USB, UART, AUDIO, GFX\}

Note that a peripheral can't initialize a write flow to write itself

\begin{center}
\resizebox{0.5\textwidth}{!}{
\includegraphics[width=.5\textwidth]{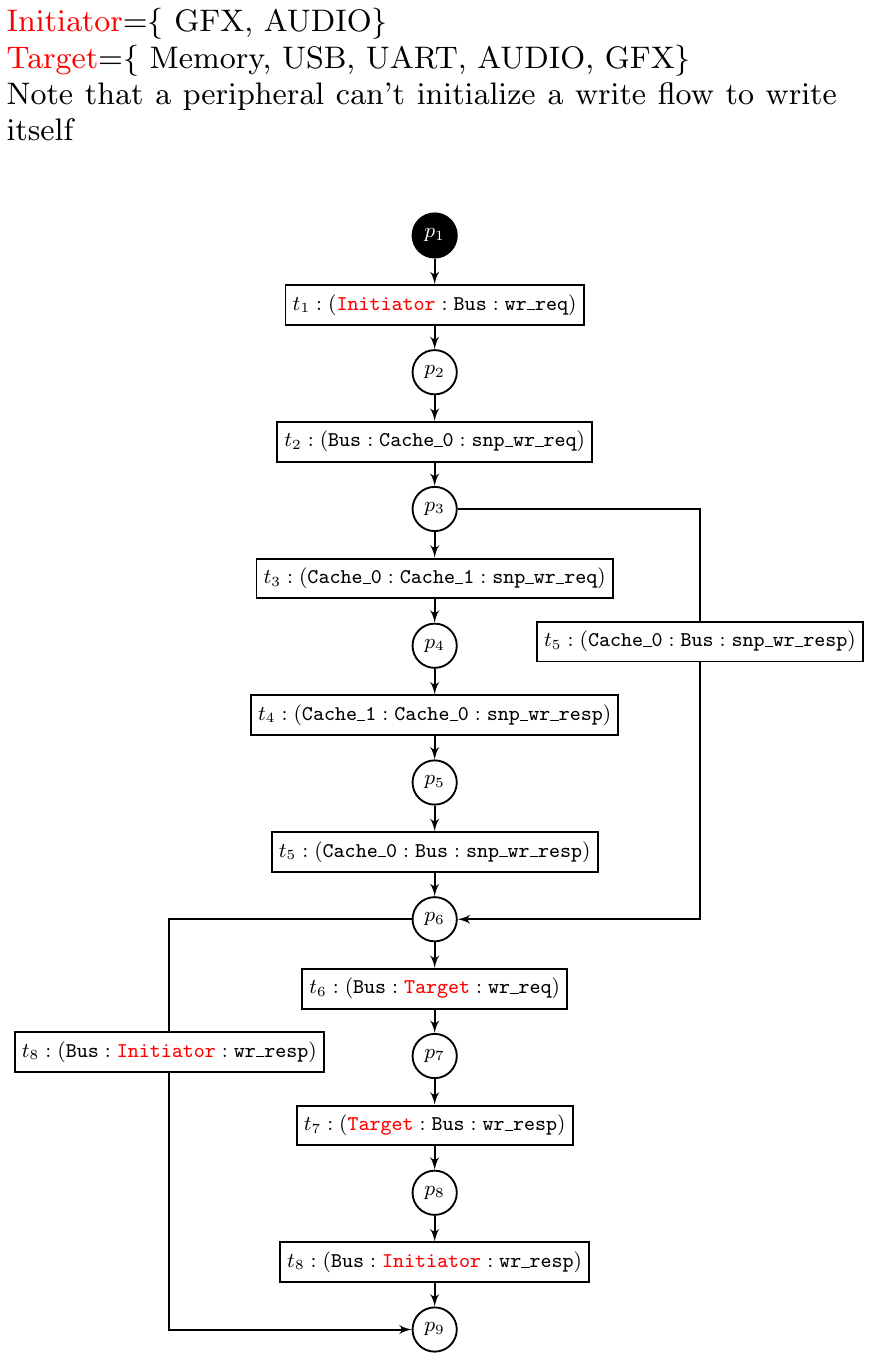}
}
\caption{LPN formalization of a upstream write protocol.
}
\label{flow-spec-uprd}
\end{center}
\end{figure}

\begin{figure}[h]
\begin{center}
\includegraphics[width=.5\textwidth]{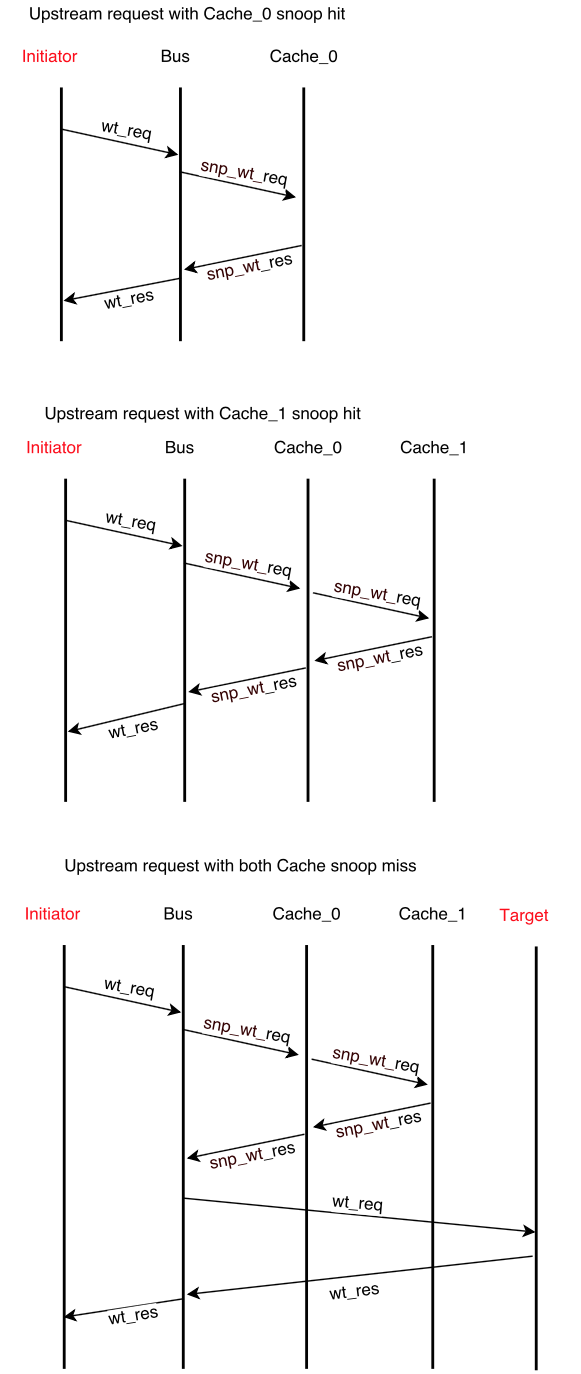}
\end{center}
\caption{MSQ of upstream write protocol}
\end{figure}%

\clearpage
\section{CPU write back protocol}
\begin{figure}[h]
{\color{red}X}=\{1,0\}

{\color{red}Target}=\{ Memory, USB, UART, AUDIO, GFX\}

\begin{center}
\resizebox{1.5in}{!}{
\includegraphics[width=0.5\textwidth]{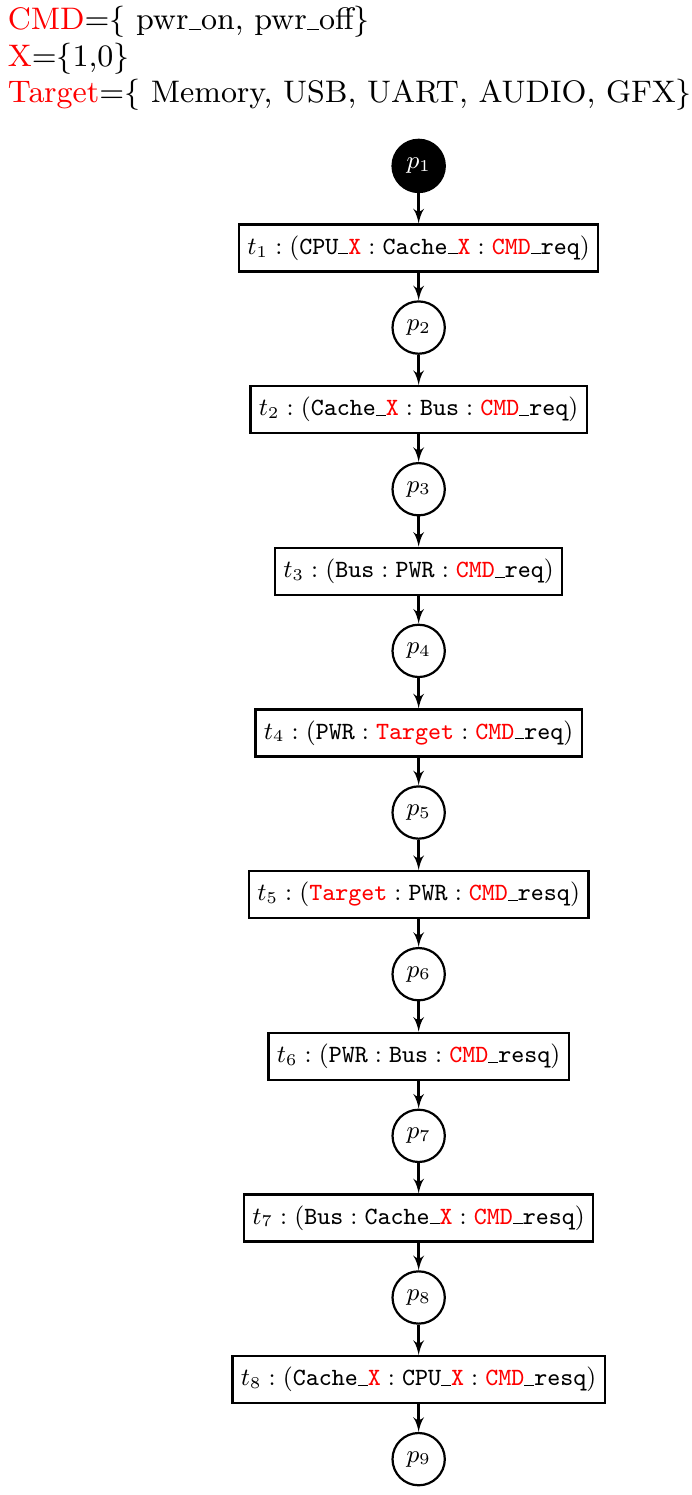}
}
\caption{LPN formalization of a CPU write back protocol.
}
\label{flow-spec-wb}
\includegraphics[width=0.5\textwidth]{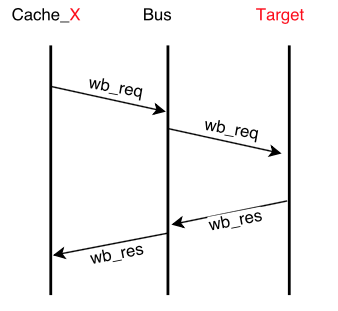}
\caption{MSQ of power write back protocol}
\end{center}
\end{figure}

\section{CPU power on/off protocol}%
\begin{figure}[h]
{\color{red}CMD}=\{ pwr\_on, pwr\_off\}

{\color{red}X}=\{1,0\}

{\color{red}Target}=\{ Memory, USB, UART, AUDIO, GFX\}
\begin{center}
\resizebox{1.6in}{!}{
\includegraphics[width=.5\textwidth]{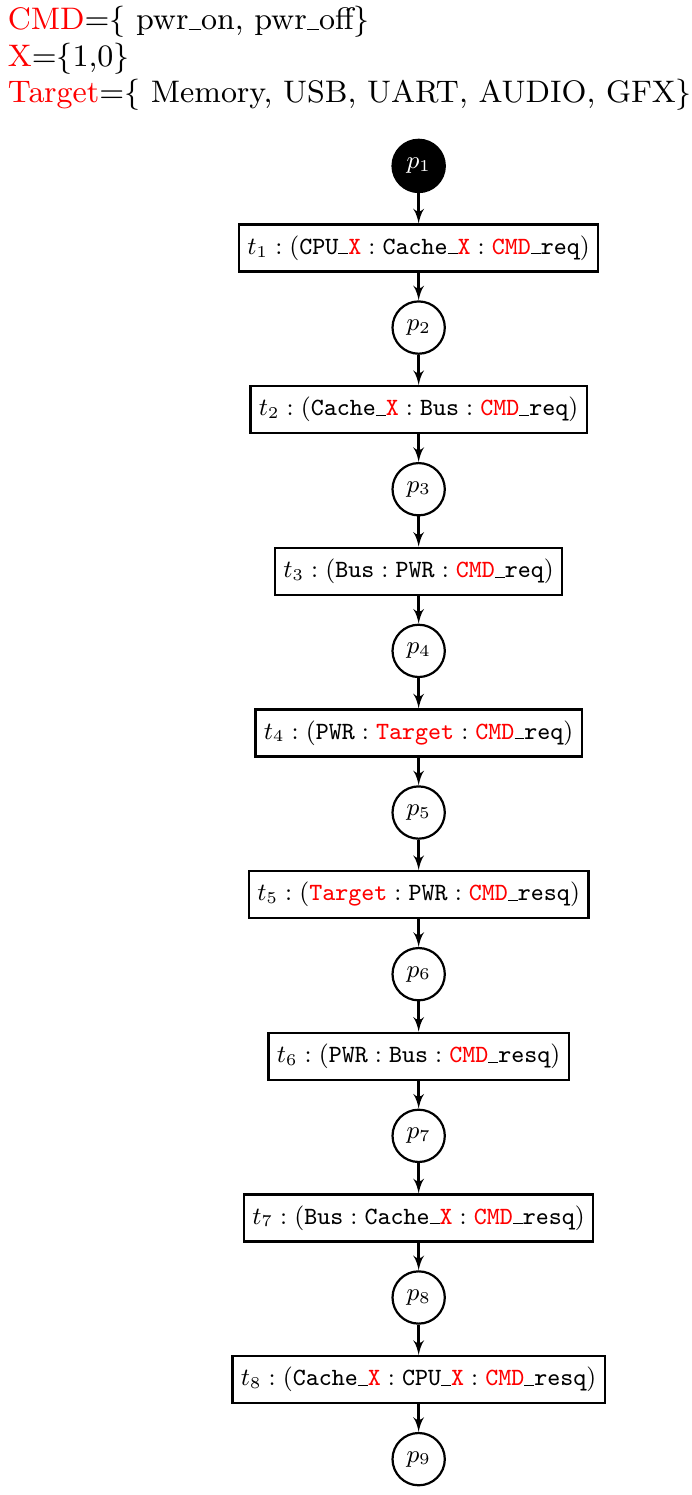}
}
\caption{LPN formalization of a CPU power on/off protocol.
}
\label{flow-spec-pwr}
\end{center}
\begin{center}
\includegraphics[width=0.4\textwidth]{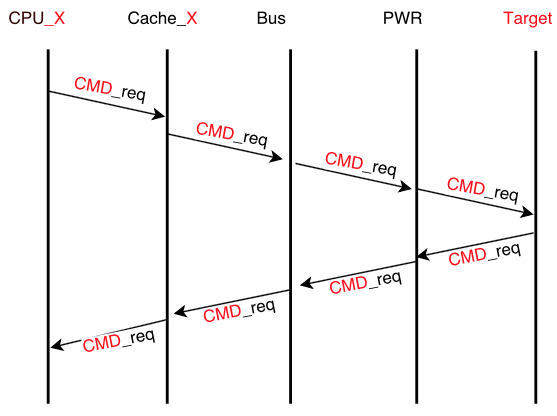}
\end{center}
\caption{MSQ of power on/off protocol}
\end{figure}



\end{document}